\documentclass[aps,prb,twocolumn,superscriptaddress]{revtex4}
\usepackage{amsmath}
\usepackage{bm}
\usepackage{graphicx}
\usepackage{amsfonts}
\usepackage{amssymb}

\begin{document}

\title{Negative differential conductance induced by electronic correlation
in a double quantum-dot molecule}
\author{Gustavo A. Lara}
\affiliation{Departamento de F\'{\i}sica, Universidad de Antofagasta, Casilla 170,
Antofagasta, Chile}
\author{Pedro A. Orellana}
\affiliation{Departamento de F\'{\i}sica, Universidad Cat\'{o}lica del Norte, Casilla
1280, Antofagasta, Chile}
\author{Enrique V. Anda}
\affiliation{Departamento de F\'{\i}sica, P. U. Cat\'{o}lica do Rio de Janeiro, C.P.
38071-970, Rio de Janeiro, RJ, Brazil}

\begin{abstract}
Electron tunneling through a two stage Kondo system constituted by a double
quantum-dot molecule side coupled to a quantum wire, under the effect of a
finite external potential is studied. We found that $I$-$V$ characteristic
shows a negative differential conductance region induced by the electronic
correlation. This phenomenon is a consequence of the properties of the two
stage Kondo regime under the effect of an external applied potential that
takes the system out of equilibrium. The problem is solved using the
mean-field finite-$U$ slave-boson formalism.
\end{abstract}

\maketitle

%\PACS{ 73.21.La, 73.63.Kv, 72.10.Fk, 85.35.Be}

\section{Introduction}

Many devices exhibit negative differential conductance ($NDC$) as multiple
quantum wells, double barrier, double quantum dots, etc \cite%
{ndc1,ndc2,ndc3,ndc4,dqd1,dqds}. The $NDC$ has applications as
amplifiers and oscillators in the microwave, mm-wave and Terahertz
frequency range. Extensive experimental and theoretical
investigation have been devoted to the study of the $I$-$V$
characteristics and $NDC$ phenomenon in double quantum dot
molecules \cite{dqd1,dqds,aguado,LOA03,fransson}. Moreover, there
is a wide literature on transport through double quantum dots
($DQD$) in different geometries, e.g, $DQD$ in series and in
parallel \cite{chi,fransson}. Two aspects of electronic transport
through quantum dots have attracted great attention in the last
years, the Coulomb blockade effect and the Kondo effect. Recently
Kondo effect has been studied in side attached \cite{THCP02} and
parallel quantum dots \cite{tanaka,sakano}. Recent electron
transport experiments showed that Kondo and Fano resonances occur
simultaneously \cite{sato}. Multiple scattering of traveling
electronic waves on a localized magnetic state are crucial for the
formation of both resonances. The condition for the Fano resonance
is the existence of two scattering channels: a discrete level and
a broad continuum band \cite{fano}.

An alternative configuration consists of a double quantum dot molecule side
coupled to a perfect quantum wire (QW)\cite{guta}. This structure is
reminiscent of the T-shaped quantum wave guides \cite{DRVRPM00}. In this
case, the QDs act as scattering centers in close analogy with the
traditional Kondo effect \cite{KCKS01}.

Although the electron-electron interaction does play an important role in
many systems that exhibit negative differential conductance, as for instance
producing bi-stabilities in the current, it is not the driving force of the
negative conductance itself. In this work instead, we study a system in the
Kondo regime with a $I$-$V$ characteristics that possesses a $NDC$ region
that is induced by the electronic correlation itself, tuned by the external
potential and the parameters of the system.

The system is constituted by a quantum dot molecule side coupled to a
quantum wire, as shown in Fig.~\ref{f:esquema}. We use the finite-$U$ slave
boson mean-field approach, which was initially developed by Kotliar and
Ruckenstein \cite{KR86} and used later by Bing Dong and X. L. Lei to study
the transport through coupled double quantum dots connected to leads \cite%
{DL01}. This approach enforces the correspondence between the impurity
fermions and the auxiliary bosons to a mean-field level to release the $%
U=\infty $ restriction. This allows to treat non-perturbatively
the dot-lead coupling for an arbitrary strength of the Coulomb
interaction $U$ \cite{DL01}.

In a previous work we study this system in a thermodynamic
equilibrium situation. We found that in the weak interaction
regime, when the direct antiferromagnetic interaction between the
dots is less than the Kondo temperature associated to the internal
dot, the transmission spectrum shows a structure with two
anti-resonances localized at the renormalized molecular energies
of the double quantum dot \cite{LOA04}. The LDOS of the system
shows that when the Kondo correlations are dominant the system is
in a two stage Kondo regime with two different temperatures
$T_{k1}, T_{k2}$, each one associated to a dot.

In the present paper we study this system under the effect of a
finite external field, which takes it out of the thermodynamic
equilibrium, modifying the Kondo regime and even destroying it,
for enough large fields. This process has fundamental consequences
on the transport properties of the system, and in particular,
creates a remarkable $NDC$ in the $I$-$V$ characteristics. This
$NDC$ phenomenon can be understood realizing that the applied bias
destroys the lower temperature Kondo state of the external dot
modifying the capability of the other to interfere on the current
that goes along the system.

\section{Model}

\begin{figure}[h]
\centering
\begin{picture}(220,100)(0,0)
    \thicklines
    \put(40,70){\line(1,0){140}}
    \put(110,70){\line(0,-1){60}}
    \put(110,10){\circle*{7}}
    \put(110,30){\circle*{7}}
    \put(100,30){\makebox(0,0){$\varepsilon_{1}$}}
    \put(100,10){\makebox(0,0){$\varepsilon_{2}$}}
    \put(120,50){\makebox(0,0){$t_{0}$}}
    \put(120,20){\makebox(0,0){$t_{c}$}}
    \put(50,75){\makebox(0,0)[b]{}}
    \put(70,75){\makebox(0,0)[b]{}}
    \put(90,75){\makebox(0,0)[b]{-1}}
    \put(110,75){\makebox(0,0)[b]{0}}
    \put(130,75){\makebox(0,0)[b]{1}}
    \put(150,75){\makebox(0,0)[b]{}}
    \put(170,75){\makebox(0,0)[b]{}}
    \put(110,90){\makebox(0,0)[b]{QW}}
    \put(150,20){\makebox(0,0)[l]{DQD}}
    \put(80,0){\dashbox{2}(60,40)}
    \put(10,70){\makebox(0,0)[r]{$\mu_{L}$}}
    \put(210,70){\makebox(0,0)[l]{$\mu_{R}$}}
    \put(40,70){\makebox(0,0)[r]{$.\, .\, .\, .\, .$}}
    \put(180,70){\makebox(0,0)[l]{$.\, .\, .\, .\, .$}}
    \multiput(50,70)(20,0){7}{\circle*{3}}
    \multiput(60,65)(20,0){6}{\makebox(0,0)[t]{}}
    \qbezier(0,50)(40,70)(0,90)
    \qbezier(220,50)(180,70)(220,90)
  \end{picture}
\caption{Scheme of double quantum dot (DQD) attached to a lead (perfect
quantum wire (QW)). The QW is coupled to the left ($L$) and right ($R$).}
\label{f:esquema}
\end{figure}
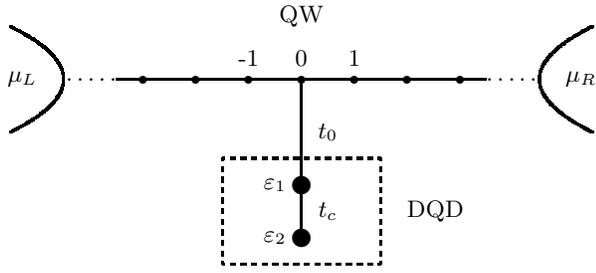

Let us consider a quantum dot molecule side coupled to a perfect
quantum wire (QW) (see Fig.~\ref{f:esquema}). We describe it by
the two-impurity Anderson Hamiltonian. Each dot has a single level
energy $\varepsilon_{l}$ (with $l = 1, 2$) and equal intra-dot
Coulomb repulsion $U$. The side attached quantum-dot molecule is
coupled to the QW with coupling $t_{0}$. The QW sites have zero
local energies and a hopping parameter $t$.

The corresponding model Hamiltonian is,

\begin{widetext}
\begin{equation}
H_{0} = -t \sum_{i,\sigma} \left( c_{i,\sigma}^{\dag} c_{i+1,
\sigma}
                             + \text{H.c.} \right)
 -\sum_{\sigma} \left[\left( t_{0} c_{0,\sigma}^{\dag}
  +t_{c} f_{2,\sigma}^{\dag} \right) f_{1,\sigma} + \text{H.c.} \right]
 +\sum_{l=1,2,\sigma}
\left[ \left( \varepsilon_{l}+\frac{U}{2}\hat{n}_{l,-\sigma}\right)
\hat{n}_{l,\sigma}
      \right] \,
\end{equation}
\end{widetext}

\noindent where $c_{i,\sigma}^{\dag}$ ($c_{i,\sigma}$) is the creation
(annihilation) operator of an electron with spin $\sigma$ at the $i$-th site
of the quantum wire; $f_{l,\sigma}^{\dag}$ ($f_{l,\sigma}$) is the creation
(annihilation) operator of an electron with spin $\sigma$ in the l-th QD, $%
\hat{n}_{l,\sigma}$ is the corresponding number operator, $t_{c}$ is the
hopping matrix element between the dots and $\varepsilon_{l}$ correspond to
the energy of the local states at the dots.

\section{Slave Boson Mean Field Theory}

To find the solution of this correlated fermions system for finite U, we
appeal to an analytical approach where, generalizing the infinite-$U$
slave-boson approximation \cite{C84} the Hilbert space is enlarged at each
site, to contain in addition to the original fermions a set of four bosons
\cite{KR86} represented by the creation (annihilation) operators $\widehat{e}%
_{l}^{\dag } $ ($\widehat{e}_{l}$), $\widehat{p}_{l,\sigma }^{\dag }$ ($%
\widehat{p}_{l,\sigma }$), and $\widehat{d}_{l}^{\dag }$ ($\widehat{d}_{l}$)
for the $l$-th dot. They act as projectors onto empty, single occupied (with
spin up and down) and doubly occupied electron states, respectively. Then,
each creation (annihilation) operator of an electron with spin $\sigma $ in
the l-th QD, is substituted by $f_{l,\sigma }^{\dag }Z_{l,\sigma }^{\dag }$ (%
$Z_{l,\sigma }f_{l,\sigma }$) where:
\begin{widetext}
\begin{equation}
Z_{l,\sigma }=\left( 1-\widehat{d}_{l}^{\dag }\widehat{d}_{l}-%
\widehat{p}_{l,\sigma }^{\dag }\widehat{p}_{l,\sigma }\right) ^{-1/2}\Bigl(%
\widehat{e}_{l}^{\dag }\widehat{p}_{l,\sigma
}+\widehat{p}_{l,-\sigma
}^{\dag }\widehat{d}_{l}\Bigr)\left( 1-\widehat{e}_{l}^{\dag }\widehat{e}%
_{l}-\widehat{p}_{l,-\sigma }^{\dag }\widehat{p}_{l,-\sigma
}\right) ^{-1/2}\,. \label{e:eq2}
\end{equation}%
\end{widetext}

\noindent As the problem is solved adopting the U-finite slave boson mean
field approximation (SBMFA), the operator is chosen to reproduce the correct
$U\rightarrow 0$ limit in the mean-field approximation without changing
neither the eigenvalues nor the eigenvector.\cite{DL01}

The constraint, i.e., the completeness relation $\sum_{\sigma }\widehat{p}%
_{l,\sigma }^{\dag }\widehat{p}_{l,\sigma }+\widehat{e}_{l}^{\dag }\widehat{e%
}_{l}+\widehat{d}_{l}^{\dag }\widehat{d}_{l}=1$ and the condition among
fermions and bosons $n_{l,\sigma }-\widehat{p}_{l,\sigma }^{\dag }\widehat{p}%
_{l,\sigma }-\widehat{d}_{l}^{\dag }\widehat{d}_{l}=0$, is
incorporated with Lagrange multipliers $\lambda _{l}^{(1)}$ and
$\lambda _{l,\sigma }^{(2)}$ into the Hamiltonian. Also in the
mean-field approximation all the boson operators are replaced by
their expectation values $p_{l,\sigma },e_{l}$ and $d_{l}$ which
can be chosen, without loss of generality, as real numbers.

The Hamiltonian in this new and enlarged Hilbert space, is, $H=H_{b}+H_{e}$,
where

\begin{eqnarray}
H_{b} &=& \sum_{l=1,2} \lambda_{l}^{(1)} \left( p_{l,\uparrow}^{2} +
p_{l,\downarrow}^{2} + e_{l}^{2} + d_{l}^{2} - 1 \right)  \notag \\
& & - \sum_{l=1,2,\sigma} \lambda_{l,\sigma}^{(2)} \left( p_{l, \sigma}^{2}
+ d_{l}^{2} \right)+ U\sum_{l=1,2} d_{l}^{2} \, ,  \label{h_boson}
\end{eqnarray}

\noindent depends explicitly only upon the boson expectation values ($%
e_{l}=\left\langle{e}_{l}\right\rangle =\left\langle{e}_{l}^{\dag}\right%
\rangle$ and equivalently for the others operators) and the Lagrange
multipliers. The Hamiltonian $H_e$ can be written,

\begin{eqnarray}
H_{e} &=&-t\sum_{i,\sigma }\left( c_{i,\sigma }^{\dag }c_{i+1,\sigma }+H.c%
\text{.}\right) +\sum_{l=1,2,\sigma }\tilde{\varepsilon}_{l,\sigma
}n_{l,\sigma }  \notag \\
&&-\sum_{\sigma }\tilde{t}_{0,\sigma }\left( c_{0,\sigma }^{\dag
}f_{1,\sigma }+H.c\text{.}\right) -\sum_{\sigma }\tilde{t}_{c,\sigma }\left(
f_{1,\sigma }^{\dag }f_{2,\sigma }+H.c\text{.}\right)  \label{h_electron}
\end{eqnarray}

\noindent The tight-binding Hamiltonian depends implicitly on the boson
expectation values through the parameters: $\tilde{\varepsilon}_{l,\sigma
}=\varepsilon _{l,\sigma }+\lambda _{l,\sigma }^{(2)}$, $\,\tilde{t}%
_{0,\sigma }=t_{0}\tilde{Z}_{l,\sigma },\tilde{t}_{c,\sigma }=t_{c}\tilde{Z}%
_{1,\sigma }\tilde{Z}_{2,\sigma }$, where $\tilde{Z}_{l,\sigma
}$is the value assumed by the operator $Z_{l,\sigma }$ when the
four boson operators are substituted by their mean values in
Equation \ref{e:eq2}.

\begin{equation}
\tilde{Z}_{l,\sigma }=\frac{p_{l,\sigma }\left( e_{l}+d_{l}\right) }{\sqrt{%
\left( 1-d_{l}^{2}-p_{l,\sigma }^{2}\right) \left( 1-e_{l}^{2}-p_{l,-\sigma
}^{2}\right) }}.
\end{equation}

The boson operator expectation values and the Lagrange multipliers
are determined by minimizing the energy $\langle \mathcal{H}
\rangle $ with respect to these quantities. It is obtained in this
way a set of nonlinear equations for each quantum dot, relating
the expectation values of the four bosonic operators, the three
Lagrange multipliers and the electronic expectation values,

\begin{subequations}
\label{e:minimizing}
\begin{align}
p_{l,\sigma }^{2}& =\langle \hat{n}_{l,\sigma }\rangle -d_{l}^{2}\,, \\
e_{l}^{2}& =1-\sum_{\sigma }\langle \hat{n}_{l,\sigma }\rangle +d_{l}^{2}\,,
\\
\lambda _{l}^{\left( 1\right) }& =\frac{t_{0}}{e_{l}}\sum_{\sigma }\langle
f_{l,\sigma }^{\dag }c_{0,\sigma }\rangle \frac{\partial \tilde{Z}_{l,\sigma
}}{\partial e_{l}}\,, \\
\lambda _{l}^{\left( 1\right) }-\lambda _{l,\sigma }^{\left( 2\right) }& =%
\frac{t_{0}}{p_{l,\sigma }}\sum_{\sigma ^{\prime }}\langle f_{l,\sigma
^{\prime }}^{\dag }c_{0,\sigma ^{\prime }}\rangle \frac{\partial \tilde{Z}%
_{l,\sigma ^{\prime }}}{\partial p_{l,\sigma }}\,, \\
U+\lambda _{l}^{\left( 1\right) }-\sum_{\sigma }\lambda _{l,\sigma }^{\left(
2\right) }& =\frac{t_{0}}{d_{l}}\sum_{\sigma }\langle f_{l,\sigma }^{\dag
}c_{0,\sigma }\rangle \frac{\partial \tilde{Z}_{l,\sigma }}{\partial d_{l}}%
\,.
\end{align}

\noindent where in the absence of external magnetic field the solutions are
spin independent.

To obtain the electronic expectation values $\langle \cdots \rangle $, the
Hamiltonian, $H_{e}$ is diagonalized. Their stationary states can be written
as
\end{subequations}
\begin{equation}
\left\vert \psi _{k}\right\rangle =\sum_{j=-\infty }^{\infty
}a_{j}^{k}\left\vert j\right\rangle +\sum_{l=1}^{2}b_{l}^{k}\left\vert
l\right\rangle \,,
\end{equation}%
where $a_{j}^{k}$ and $b_{l}^{k}$ are the probability amplitudes
to find
the electron at the site $j$ and at the $l$-th QD respectively, with energy $%
\omega =-2t\cos k$. As we study the paramagnetic case the spin index is
neglected.

The amplitudes $a_{j}^{k}$ and $b_{l}^{k}$ obey the following linear
difference equations

\begin{subequations}
\label{e:diferencias}
\begin{align}
\omega a_{j}^{k}& =-t(a_{j+1}^{k}+a_{j-1}^{k})\,,\quad j\neq 0\,, \\
\omega a_{0}^{k}& =-t(a_{1}^{k}+a_{-1}^{k})-\tilde{t}_{0}b_{1}^{k}\,, \\
(\omega -\tilde{\varepsilon}_{1})b_{1}^{k}& =-\tilde{t}_{0}a_{0}^{k}\,-%
\tilde{t}_{c}b_{2}^{k}, \\
(\omega -\tilde{\varepsilon}_{2})b_{2}^{k}& =-\tilde{t}_{c}b_{1}^{k}\,.
\end{align}

In order to study the solutions of Eqs.~\eqref{e:diferencias}, we
assume that the electrons are described by an incident, a
reflected and a transmitted plane waves with unitary, $r$ and
$\tau $ amplitudes, respectively.\cite{PPE} That is,

\end{subequations}
\begin{subequations}
\label{e:solut}
\begin{align}
a_{j}^{k}& =e^{\text{i}k\cdot j}+re^{-\text{i}k\cdot j}\,;\,\,\left( k\cdot
j<0\right) \,, \\
a_{j}^{k}& =\tau e^{\text{i}k^{\prime }\cdot j}\qquad \,;\qquad \left(
k\cdot j>0\right) .
\end{align}

Inserting Eqs.~\eqref{e:solut} into Eqs.~\eqref{e:diferencias}, we get an
inhomogeneous system of linear equations for $\tau $, $r$, $a_{j}^{k}$ and $%
b_{l}^{k}$, leading to the following expression in equilibrium ($k=k^{\prime
}$)
\end{subequations}
\begin{equation}
\tau =\frac{(\omega -\tilde{\varepsilon}_{-})(\omega -\tilde{\varepsilon}%
_{+})}{(\omega -\tilde{\varepsilon}_{-})(\omega -\tilde{\varepsilon}_{+})+%
\text{i}(\omega -\tilde{\varepsilon}_{d2})\tilde{\Gamma}}\,,
\label{t-amplitude}
\end{equation}%
where the bonding ($\tilde{\varepsilon}_{-}$) and antibonding energy ($%
\tilde{\varepsilon}_{+}$) are defined by $\tilde{\varepsilon}_{\pm }=(\tilde{%
\varepsilon}_{d1}+\tilde{\varepsilon}_{d2})/2\pm \sqrt{(\tilde{\varepsilon}%
_{d1}+\tilde{\varepsilon}_{d2})/2)^{2}+\tilde{t}_{c}^{2}}$ and $\tilde{\Gamma%
}=\pi {\tilde{t}}_{0}^{2}\rho _{0}$ is the renormalized coupling
between the double quantum-dot and the quantum wire and $\rho
_{0}$ is the density of states of the leads at the Fermi level. In
spite of the apparent simplicity
of the expression, it is necessary to remember that the quantities $\tilde{t}%
_{0}$ and $\tilde{t}%
_{c}$ implicitly depend on the expectation values of the boson and
fermion operators.

\bigskip

The transmission probability is given by $T=|\tau |^{2}$,

\noindent\ \
\begin{equation}
T(\omega )=\frac{[(\omega -\tilde{\varepsilon}_{-})(\omega -\tilde{%
\varepsilon}_{+})]^{2}}{[(\omega -\tilde{\varepsilon}_{-})(\omega -\tilde{%
\varepsilon}_{+})]^{2}+[(\omega -\tilde{\varepsilon}_{d2})\tilde{\Gamma}]^{2}%
}\,.
\end{equation}

From the amplitudes $b_{1}^{k}$ and $b_{2}^{k}$ we obtain the local density
of states (LDOS) at the quantum dot $l$ (with $l=1,2$).
\begin{align}
\rho _{1}& =\frac{1}{\pi }\frac{\tilde{\Gamma}(\omega -\tilde{\varepsilon}%
_{d2})^{2}}{[(\omega -\tilde{\varepsilon}_{-})(\omega -\tilde{\varepsilon}%
_{+})]^{2}+[(\omega -\tilde{\varepsilon}_{d2})\tilde{\Gamma}]^{2}}\,, \\
\rho _{2}& =\frac{1}{\pi }\frac{\tilde{\Gamma}\tilde{t}_{c}^{2}}{[(\omega -%
\tilde{\varepsilon}_{-})(\omega -\tilde{\varepsilon}_{+})]^{2}+[(\omega -%
\tilde{\varepsilon}_{d2})\tilde{\Gamma}]^{2}}\,.
\end{align}

\bigskip

In the nonequilibrium case, we suppose a finite source-drain
biased with a symmetric voltage drop. The incident electrons from
the left side ($L$) are in equilibrium with the thermodynamical
potential $\mu _{L}=\mu +V/2$ and those from the right side ($R$)
with $\mu _{R}=\mu -V/2.$

Once the amplitudes $a_{j,\sigma }^{k}$ and $b_{j,\sigma }^{k}$ are known,
the electronic expectation values are obtained from,

\begin{equation}
\langle f_{l}^{\dag }c_{j}\rangle =\frac{1}{2}\sum_{\alpha =L,R}\frac{1}{N}%
\sum_{k_{\alpha }}f\left( \epsilon _{k_{\alpha }}-\mu _{\alpha }\right)
b_{l}^{k_{\alpha }\ast }a_{j}^{k_{\alpha }},
\end{equation}

where $\epsilon _{k_{\alpha }}=-2t\cos k_{\alpha }.$ The current given by,
\begin{equation}
I=2\frac{2e}{\hbar }t\sum_{\alpha ,k_{\alpha }}f\left(
\epsilon_{k_{\alpha }}-\mu _{\alpha }\right)
{Im}\{a_{0}^{k_{\alpha }}a_{1}^{k_{\alpha }}\}
\end{equation}

\noindent where $f\left( \epsilon _{k_{\alpha }}-\mu _{\alpha }\right) $ it
is the Fermi function for incident electrons from the $\alpha $ side and the
sum on $k_{\alpha }$ is taken up to the maximum value $\cos ^{-1}(-\mu
_{\alpha }/2t).$

\begin{figure}[h]
\centering
\includegraphics[angle=0, scale=0.3]{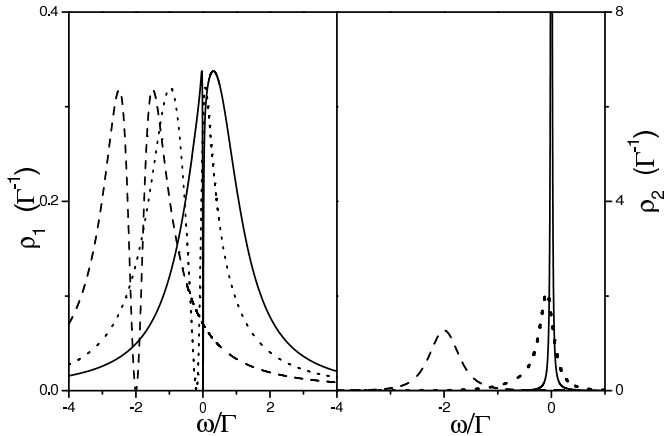}
\caption{LDOS at each dot,left panel QD1, right panel QD2, for $%
t_c=0.5\Gamma $ (solid line), $V_g=-2\Gamma$ for various values of on site
energy, $U=6\Gamma $(solid line), $U=2\Gamma$(dotted line) $U=0$(dashed
line) }
\label{f:fig2}
\end{figure}

\begin{figure}[t]
\centering
\includegraphics[angle=0, scale=0.3]{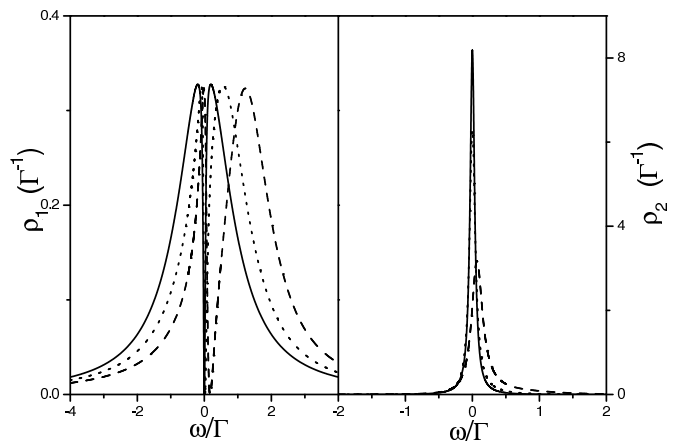}
\caption{LDOS at each dot, left panel QD1, right panel QD2, for $%
t_c=0.5\Gamma$ (solid line), $U=4\Gamma$ for various values of the gate
voltage,$V_g=-2\Gamma$(solid line),$V_g=-\Gamma$(dotted line) and $V_g=0$%
(dashed line)}
\label{f:fig3}
\end{figure}

The quantities $\varepsilon _{l}$, the energies of the local states at the
dots, are taken to be equal to $\varepsilon _{1}=\varepsilon _{2}=V_{g}$,
where $V_{g}$ is the gate voltage applied to the quantum dots.

\section{Results}

\subsection{Thermodynamics Equilibrium Case}

In order to obtain a more clear insight, we study first the system
in thermodynamical equilibrium. The DOS at each dot of the
quantum-molecule is shown in Figure \ref{f:fig2}, for various
values of $U$,
for the case $t_{c}=0.5\,\Gamma $ and $V_{g}=-2\Gamma $. As U increases, $%
U>2\Gamma $, the system passes from the intermediate valence
regime into a Kondo regime. This process is clearly seen to occur
for QD2 where the resonance shifts and becomes sharper as the
system enters into the Kondo regime. The same process takes place
for the QD1 although the sharpening is less accentuated. It is
noticeable that as QD2 develops a Kondo resonance,  a dip appears
at the Fermi energy in the LDOS of QD1. This is produced because
the spin of the local electron at QD2 is Kondo correlated with the
conduction band spins through the mediation of the intermediate
dot, that is as well at the Kondo regime. This coupling creates a
sharp peak at the LDOS of QD2 with width $T_{k2}$ and a depletion
of this same width at the Fermi level of the Kondo peak of width
$T_{k1}$ corresponding to QD1. This result implies the existence
of  a two stage Kondo regime that appears in this system in the
weak interacting limit, when $t_{c}^{2}/U<T_{k1}$ \cite{LOA04},
being $T_{k1}$ and $T_{k2}$ the Kondo temperatures associated to
each dot. It is the depletion of the LDOS of QD1 at the Fermi
level that permits the transmission to be unity. In this case,
although the side attached QD1, is in the Kondo regime, it does
not provide an alternative path for the conducting electrons. As a
consequence, there is no destructive interference because there is
only the direct path available for them.

In Figure \ref{f:fig3} we show the LDOS of each QD for different values of $%
V_{g}$ in the two stage Kondo regime. As expected, the peaks are
pinned at the Fermi energy independent of the values of the
$V_{g}$, a clear signal of the Kondo regime. Here it is clearly
seen at the Fermi energy, the antiresonance at it QD2 and the
concomitant resonance at the QD1.

\begin{figure}[h]
\centering
\includegraphics[angle=0, scale=0.3]{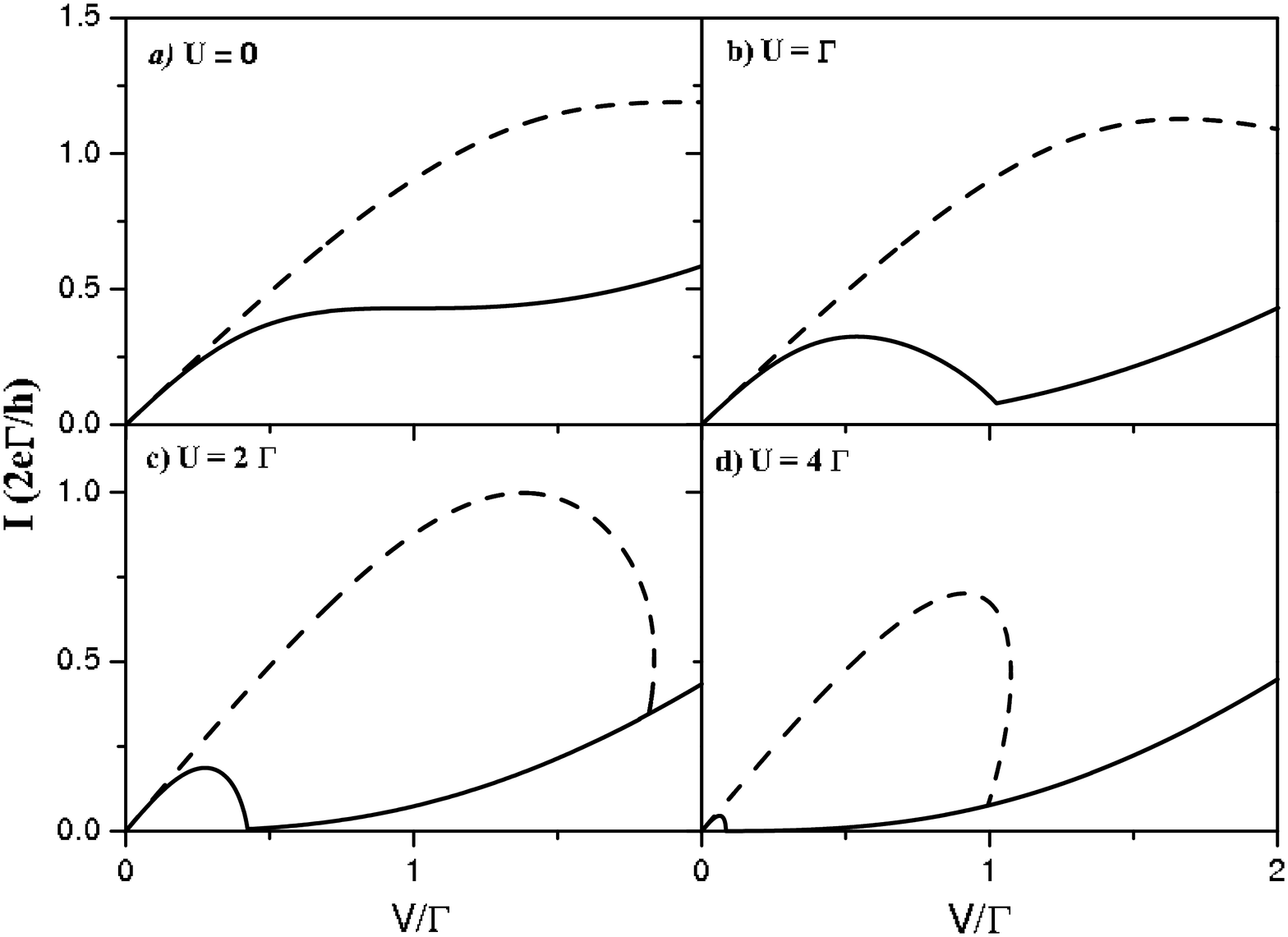}
\caption{$I$-$V$ characteristics for $t_c=0.5\Gamma$ (solid line) and $%
t_c=\Gamma$ (dashed line) for various values of $U$ with $\protect\varepsilon%
_{1}= \protect\varepsilon_{2}=-U/2$}
\label{f:fig4}
\end{figure}

\subsection{The Out of Thermodynamical Equilibrium Situation}

The previous discussion was restricted to the thermodynamical
equilibrium situation and was presented with the purpose of
clarifying the concepts involved. When an external potential is
applied the scenario changes completely. The physics of this new
situation can be explained analyzing the current and the
differential conductance $dI/dV$, two significant and
experimentally measurable quantities, as a function of the applied
field.

Figure \ref{f:fig4} displays the $I$-$V$ characteristics (solid
line) for two values of $t_c $ and different values of $U$, with
$V_g=-U/2$. For $U=0$, the $I$-$V$ characteristics shows a plateau
that is related to the Fano antiresonances in the transmission
spectrum. When the applied potential is of the order of the
interdot interaction, the transmission, due to the Fano
destructive interference, is almost zero in the bonding and
antibonding regions giving no additional contribution to the total
current as the applied potential is increased. This is the origin
of the plateau behavior, shown in the Figure \ref{f:fig4}, when
U=0. As $U$ is increased, a negative differential conductance
appears in the $I$-$V$ characteristics that gets more important as
$t_c$ is augmented.
\begin{figure}[h]
\centering
\includegraphics[angle=0, scale=0.25]{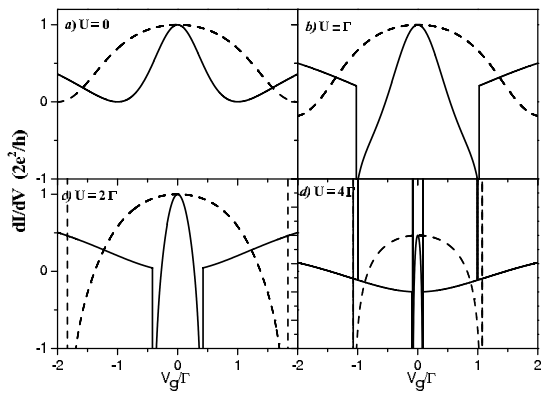}
\caption{Differential conductance for $t_{c}=0.5\Gamma$ (solid line) and $%
t_{c}=\Gamma$ (dashed line), for various values of $U$ with $V_g=-U/2$}
\label{f:fig5}
\end{figure}

Figure \ref{f:fig5} depicts the differential conductance for the
same parameters of Fig.\ref{f:fig4}. For $U=0$ it reflects
essentially the transmission spectrum. As $U$ is increased the
differential conductance becomes negative in a region of the
applied potential, reflecting the fact that the $NDC$ is a
consequence of the
Coulomb interaction. It is a small effect in the fluctuating valence regime $%
\Gamma >U$ and develops completely in the Kondo regime $\Gamma <U$,
increasing with $t_{c}.$

In order to get insight into these results it is convenient to
write the transmission (Eq.11) as the superposition of a Fano and
a Breit-Wigner line shapes, a good approximate expression for
large values of $U$\cite{LOA04}. The results is,

\begin{equation}
T(\omega )\approx \frac{\epsilon ^{2}}{\epsilon ^{2}+1}+\frac{{\tilde{\Delta}%
}^{2}}{\omega ^{2}+{\tilde{\Delta}}^{2}}\,,  \label{e:transmission}
\end{equation}%
where $\epsilon =\omega /\tilde{\Gamma}$ and ${\tilde{\Delta}}=\tilde{t}%
_{c}^{2}/\tilde{\Gamma}$ corresponds to the two Kondo temperatures $T_{k1}=%
\widetilde{\Gamma }$ and $T_{k2}={\tilde{\Delta}.}$

An analytical expression can be obtain  for the current by integrating over $%
\omega$ the transmission probability given in
Eq.~\eqref{e:transmission}.

\begin{equation}
I \approx \frac{2e}{h}\left[ eV-2\tilde{\Gamma}\arctan \left( \frac{eV}{%
\tilde{2\Gamma}}\right) +2\tilde{\Delta}\arctan \left( \frac{eV}{2\tilde{%
\Delta}}\right) \right] ,  \label{current}
\end{equation}

We identify each term of the Eq.\eqref{current} as follows. The
first term in the right side is the contribution arising from an
ideal one dimensional conductor. The second term arises from the
Kondo-Fano state with temperature $T_{k1}$ giving rise to a quasi
plateau for the current and almost zero differential conductance
when $|V|\ll \tilde{\Gamma}$. The third term arises from the Kondo
state weakly coupled to the wire and it is responsible for the
rapid increase of the current in the region of small applied
potentials.

\begin{figure}[t]
\centering
\includegraphics[angle=0, scale=0.25]{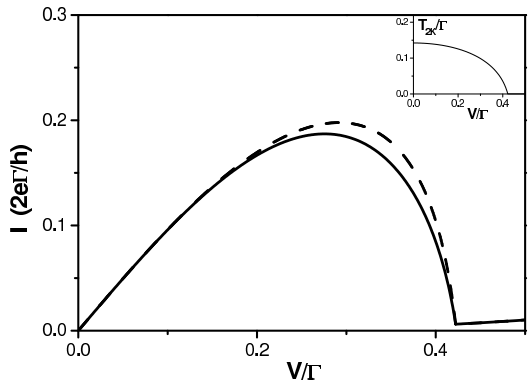}
\caption{Comparison of $I-$$V$ characteristics between the numerical
calculation (solid line)and the approximation (Eq.4)(dashed line) for $%
U=2\Gamma$, $t_c=0.5\Gamma$ with $V_{g}=-U/2$. $T_{2K}=\tilde{\Delta}$ in
the inset }
\label{f:fig6}
\end{figure}

It is important to emphasize that in this expression the quantities $\tilde{%
\Delta}$ and $\tilde{\Gamma}$ are functions of $V$ obtained through the
self-consistent calculation presented above. The case of $\tilde{\Delta}%
=T_{k2}$ is shown in the inset of Fig.\ref{f:fig6}. It is clear
that increasing $V$, both the Kondo temperature of the external
dot and the current reduce to zero simultaneously with the
disappearance of the $NDC$ phenomenon. As mentioned above, this
behavior can be understood realizing that the effect of the
external quantum dot is to reduce the intermediate dot
interference effect on the current circulating along the leads.
This role is exercised by the external quantum dot as far as it is
at the Kondo regime. When the external potential is large enough
as to disrupt its Kondo ground state ($T_{k2}<V$)the interference
 is reestablished and the current goes to
zero. As $T_{k2}$ increases with $t_{c}$, for larger values of
$t_{c}$ this disrupting process requires bigger values of the
applied potential as shown in Fig.\ref{f:fig4}. When
$|V|$increases still further it destroys as well the Kondo regime
of the intermediate quantum dot and the current rises as depicted
in this same figure. This seems to be the behavior of a two stage
Kondo system under the effect of an external potential. The Kondo
regime of the outside quantum dot, not directly connected to the
continuum, depends upon the Kondo effect of the intermediate
quantum dot. As its Kondo temperature is lower, its Kondo ground
state is disrupted by lower values of $V$ than the other dot. This
process manifests in the transport properties by the appearance of
a $NDC$ region in the current. The Fig.\ref{f:fig6} displays a
comparison between the $I$-$V$ characteristics of the numerical
calculation and the approximation (Eq. (17)) for $U=2\Gamma $ and $%
t_{c}=0.5\Gamma $. The approximation over estimates the peak of the current
however, qualitatively, it maintains the form.

Deriving the current in Eq.(\ref{current}) we obtain
\begin{widetext}
\begin{equation}
\frac{\partial I}{\partial V}\approx \frac{2e^{2}}{h}\left\{ %
\frac{\left( \frac{eV}{2\tilde{\Gamma}} \right) ^2 }          %
     {1+\left( \frac{eV}{2\tilde{\Gamma}} \right) ^2 }        %
+\frac{1}{1+\left( \frac{eV}{2\tilde{\Delta}} \right) ^2 }    %
+2\frac{\partial \tilde{\Delta} }{\partial V}               %
\left[ \arctan \left( \frac{eV}{2\tilde{\Delta}}\right) %
 -\frac{\left( \frac{eV}{2\tilde{\Delta}}\right)}           %
    {1+\left( \frac{eV}{2\tilde{\Delta}}\right) ^2}        %
\right] \right\} .
\end{equation}
\end{widetext}

In the above equation, the last term is responsible for the
negative
differential conductance. This term is negative because $\partial \tilde{%
\Delta} / \partial V<0$. This expression is able to reproduce very well the
results of the differential conductance shown in Fig.\ref{f:fig5}.

Regarding the observability of the $NDC$ in the side-coupled double quantum
dot molecule, we consider the value of $\Gamma$ given by Sato et al. \cite%
{sato}, $\Gamma=3 meV$. For $tc=0.5\Gamma$ and $U=\Gamma$ the
maximum and minimum of the current are $I_{max} \sim 24$ nA and
$I_{min} \sim 6$ nA, respectively, giving a peak to valley ratio
of 4:1. The lower Kondo temperature of the external dot for these
same parameters is of the order of 2 K. These values of the
current and temperature are well above the experimental limits of
present day techniques.

\section{Summary}

In summary we have studied the nonequilibrium transport through a
double quantum dot molecule side-coupled to a quantum wire using
the finite-$U$ slave boson mean field approach at $T=0$ as a
function of the parameters that define the system. We find that
the $I$-$V$ characteristics shows a remarkable $NDC$, different
from case reported in the literature, is induced by the electronic
correlation. This $NDC$ behavior is a consequence of the
properties of a two stage Kondo system under the effect of an
external applied potential that takes the system out of
equilibrium when the applied potential is large enough to destroy
the Kondo regime characterized by the lower Kondo temperature.

\section*{Acknowledgement}

We would like to thank M.A. Davidovich for her very useful
comments of the manuscript. This work was partially financial
supported by CONICYT/Programa Bicentenario de Ciencia y Tecnologia
(CENAVA, grant ACT27), CIAM (CONICYT-CNPq-NSF) and FONDECYT under
grant 1040385. G.A.L. thank U.A. (PEI-1305-04). E.V.A.
acknowledges support from the Brazilian agencies, CNPq and FAPERJ.
%%%%%%%%%%%%%%%%%%%%%%%%%%%%%%%%%%%%%%%%%%%%%%%%%%%%%%%%%%%%%%%%%%%%%

\end{document}